\documentclass[aps,prl,superscriptaddress,rsi,amsmath,amssymb,
 reprint,
 floatfix,
]{revtex4-2}
\usepackage{graphicx}
\usepackage{dcolumn}
\usepackage{bm}
\usepackage{comment}
\usepackage{lipsum}
\usepackage{textcomp}
\usepackage{gensymb}
\usepackage{color,colordvi}
\usepackage[usenames,dvipsnames]{xcolor}

\begin{document}

\title{Predicting structure and swelling of microgels with different crosslinker concentrations combining machine-learning with numerical simulations}
\date{\today}

\author{Susana Mar\'in-Aguilar}
\email{susana.marinaguilar@uniroma1.it}
\affiliation{Department of Physics, Sapienza University of Rome, Piazzale Aldo Moro 2, 00185 Roma, Italy}
\author{Emanuela Zaccarelli}
\email{emanuela.zaccarelli@cnr.it}

\affiliation{Department of Physics, Sapienza University of Rome, Piazzale Aldo Moro 2, 00185 Roma, Italy}
\affiliation{CNR Institute of Complex Systems, Uos Sapienza, Piazzale Aldo Moro 2, 00185, Roma, Italy}

\begin{abstract}

Microgels made of poly(N-isopropylacrylamide) are the prototype of soft, thermoresponsive particles widely used to study fundamental problems in condensed matter physics. However, their internal structure is far from homogeneous, and existing mean-field approaches, such as Flory-Rehner theory, provide only qualitative descriptions of their thermoresponsive behavior. Here, we combine machine learning and numerical simulations to accurately predict the concentration and spatial distribution of crosslinkers, the latter hitherto unknown experimentally, as well as the full swelling behavior of microgels, using only polymer density profiles. Our approach provides unprecedented insight into structural and thermodynamic properties of any standard microgel, including experimental ones. 
\end{abstract}

\maketitle

Microgels are colloidal particles resulting from a complex chemical synthesis, that are nowadays used as the simplest model system of soft, deformable particles~\cite{karg2019nanogels,scotti2022softness}. Since they are entirely made of polymers, arranged in a disordered network via chemical crosslinking, they are able to respond to external stimuli. For example, they can easily change their volume in response to a temperature variation when the constituent polymer is thermoresponsive, such as for poly(N-isopropylacrylamide) (pNIPAM)~\cite{otten2024volume,del2021two,conley2016superresolution}. Therefore, at a characteristic temperature, the polymer network undergoes a so-called Volume Phase Transition (VPT), from a swollen to a collapsed state. This can be exploited to study fundamental physics problems via an in-situ tuning of the suspension packing fraction~\cite{schelling2024controlled,appel2016mechanics}. One important parameter that can be easily varied experimentally is the molar concentration of crosslinking molecules, hereafter abbreviated as crosslinker concentration $c$, which controls the overall softness of the microgel~\cite{scotti2022softness}. It is well-known that, in the \textit{standard} synthesis protocol --- precipitation polymerization --- crosslinkers react faster than monomers, thereby accumulating within a central region of the microgel, the so-called core, which can be distinguished from the outer corona, where crosslinkers are almost absent. Despite several attempts using chemical kinetic modeling, to our knowledge, there is no available work able to predict the crosslinker distribution within pNIPAM microgels.
Instead, to describe the VPT, there exists the celebrated theory of Flory and Rehner~\cite{flory1953principles}, which combines mean-field-like theoretical assumptions with more phenomenological approaches, to describe the available experimental swelling curves. However, it is well-known that the theory suffers some important drawbacks, such as the presence of many fit parameters, whose physical meaning is often obscure~\cite{lopez2017does}.  Therefore, it would be important to have an alternative way to predict the swelling behavior of a microgel. 

In this Letter, we take advantage of a recently put forward  monomer-resolved microgel model, that has been shown to accurately describe the internal structure of the experimental systems, both in bulk suspensions~\cite{ninarello2019modeling}
and at liquid-liquid interfaces~\cite{camerin2019microgels}, to fill this gap.
To this aim, we perform extensive computer simulations of individual microgels with different crosslinker concentrations to create a comprehensive numerical database of microgel structures at many different temperatures across the VPT. Using a subset of these structures as training set, we apply machine learning (ML) techniques to predict crosslinker properties. Building on the concept that density profiles encode all relevant structural information, we demonstrate that the total polymer density profiles of microgels in the swollen state are enough to accurately predict the crosslinker concentration via unsupervised ML. This approach is subsequently validated against simulations of microgels with different sizes and $c$ outside the training set, as well as for available experimental data. Following a similar approach, we then train a supervised neural network (NN) to predict the distribution of the crosslinkers, which is found to follow the fuzzy sphere model~\cite{stieger2004small,bergmann2018super}, in analogy to the overall microgel. Our approach thus enables, for the first time, the prediction of the crosslinker distribution in experimental systems. Finally, we use our database to build a phenomenological framework to predict the complete swelling behavior of the microgels for a given $c$, similar in spirit to the Flory-Rehner theory. 
Therefore, from the sole knowledge of the microgel total density profile at low temperature, which can be extracted from the fit of the form factor, routinely measured nowadays by small-angle scattering techniques~\cite{stieger2004small}, or directly obtained by super-resolution microscopy~\cite{bergmann2018super}, our ML approach is able to reliably predict the  concentration of crosslinkers and their radial distribution, as well as the full swelling behavior of any standard pNIPAM microgel.

\begin{figure*}[th!]
\begin{center}
\includegraphics[width=1.0\linewidth]{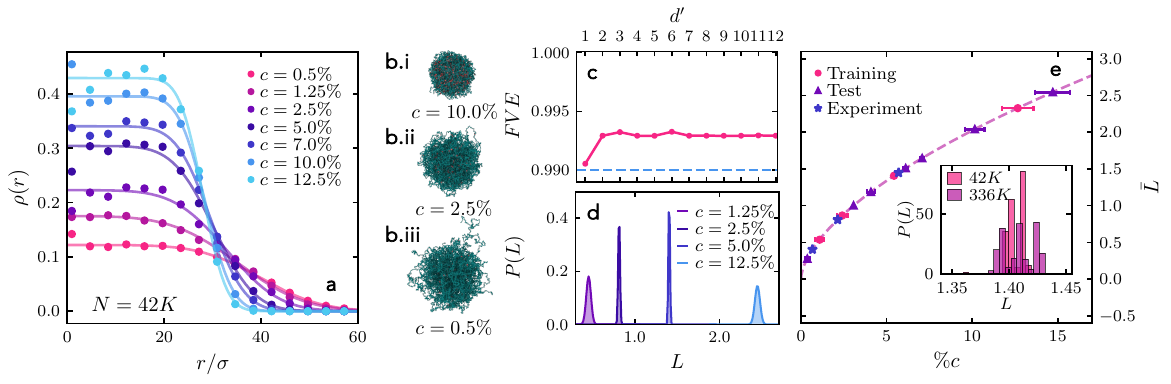}
\end{center}
\vspace{-0.5cm}
\caption{
\textbf{Crosslinker concentration prediction.} a) Averaged radial density profiles of microgels with $N\sim 42000$ for various crosslinker molar fractions $c$ from simulations (symbols) and  fuzzy sphere fits (lines); b) snapshots of high (i), medium (ii) and low (iii) crosslinked microgels;  c) Fraction of Variance Explained (FVE) for the autoencoder with varying dimension $d'$ of the latent space $L$. The horizontal line indicates $FVE=99\%$;
d) probability distributions of $L$ for $N\sim 336000$ microgels with different $c$ for the training data; e) $L$ as a function of $c$ for all simulated microgels, including training (circles), test (triangles) data sets and experiments from Ref.~\cite{hazra2023structure}, corresponding to microgels of $c=1.25,2.5,5.0$\% (stars). The dashed line is a power-law fit. Inset: $P(L)$ for 
two different microgel sizes with $c=10.0$\%.
}
\label{fig:ae_dens}
\end{figure*}

In order to generate a database of microgel particles we perform molecular dynamics (MD) simulations of monomer-resolved microgels, interacting with the bead-spring potential~\cite{kremer1990dynamics,soddemann2001generic}, as described in the End Matter (EM). Microgels are made of a total number of beads either $N\sim 42000$ or $N\sim 336000$, of which a fraction $c$ identifies the molar crosslinker concentration varying from  $c=0.5$\% to $c=15$\%, covering the usually employed experimental range.
We start by reporting the dependence of microgel structure on crosslinker concentration in the swollen state. Fig. \ref{fig:ae_dens}(a) shows the radial density profiles $\rho(r)$ of the microgels with $N\sim 42000$. The change in $c$ has a direct effect on the shape of the microgel, as reflected in the profiles, which become progressively sharper with increasing $c$. In particular, the behavior of $\rho(r)$ displays an almost constant regime, representing the core, followed by a decay at large $r$ in the corona region. The latter becomes progressively more extended and less distinct from the core with decreasing $c$. Typical snapshots of the microgels are reported in Fig.~\ref{fig:ae_dens}(b), changing from a rather compact structure for $c=10$\%  to a much more heterogeneous one for $c=0.5$\%, denoting the presence of so-called dangling ends in the exterior of the microgel. All density profiles are well-described by the fuzzy sphere model~\cite{stieger2004small}, as shown in Fig.~\ref{fig:ae_dens}(a), with the corresponding fit parameters reported in Table S1 of the Supplemental Material (SM)~\cite{sm}. 

Variations in $c$ also affect the microgel size. For instance, the hydrodynamic radius $R_H$ displays a power-law dependence $R_H \sim c^{-0.21\pm0.01}$, as shown in the SM~\cite{sm} (Fig.~S1), in good agreement with the predictions of the Flory-Rehner theory for polymer networks~\cite{flory1953principles,paulin2021revisiting}. This behavior, also shared by the calculated gyration radius, was also previously reported in experimental works~\cite{fernandez2009gels,hazra2023structure}. Instead, at the local scale, for individual polymer chains belonging to the network, the average end-to-end distance varies as $R_{ee} \sim c^{-0.54\pm0.01}$ (see Fig.~S1). This behavior is relatively close to the Flory scaling for polymer chains in good solvent~\cite{hazra2023structure}, with a minor deviation likely attributed to the effect of the crosslinkers.

The previous results suggest that the knowledge of the crosslinker concentration is encoded in some of the microgel features, in particular in the total density profile of the microgels. 
Therefore, in order to predict $c$ of a given microgel, we propose to use unsupervised ML by means of autoencoders (AE) to extract this information from $\rho(r)$. This method consists of a neural network that encodes a high-dimensional input and projects it into a low-dimensional latent space. The latter is then decoded recovering the initial input as closely as possible. The key information that characterizes the initial input data is presumably conserved in the latent space, and hence, we expect this to correspond to the crosslinker concentration. 

To train the AE, we employ discretized density profiles obtained from simulations of microgels with varying $c$. In particular, the input of the AE, is a vector $\mathbf{P}$ of dimension $d=115$, where each entry corresponds to the density profile value at a given radial distance. To make the method scalable and transferable to microgels of different sizes, we preprocess the input data by rescaling the radial coordinate as described in the EM. This allows the method to be applied to any microgel, whose density profile is known, including experimental ones. Further, to provide sufficient data and variability to the AE, we employ averaged radial density profiles in small windows of time. We use $c=1.25, 2.5, 5.0, 12.5$\%  as a training set, while the simulations for additional $c$ values are later used as test data sets.  
The AE consists of a neural network of $2$ layers, the first one corresponding to the encoder with a hidden dimension of $80$, followed by a bottleneck that projects the data onto a $d'$-dimensional space, and finally, the decoder that recovers the input data. 
We first determine the minimum number of dimensions to be used in the bottleneck $d'$. 
To this aim, we train the AE for at least $3\times 10^3$ epocs varying $d'$ and then we compute the final mean squared error (MSE) and the fraction of variance explained (FVE) as defined in the EM. The resulting FVE is reported  in Fig.~\ref{fig:ae_dens}(c) as a function of $d'$, showing that, even for $d'=1$, the AE is already able to capture more than $99\%$ of the variance of the input data by projecting it onto a single number. At higher dimensions, $d'\geq 2$, this improves, reaching a plateau of $FVE\approx0.993$. However, for simplicity, we choose to work with a bottleneck dimension of $d'=1$ for the prediction of $c$, since it already yields very accurate results. 

\begin{figure}[th!]
\begin{center}
\includegraphics[width=0.85\linewidth]{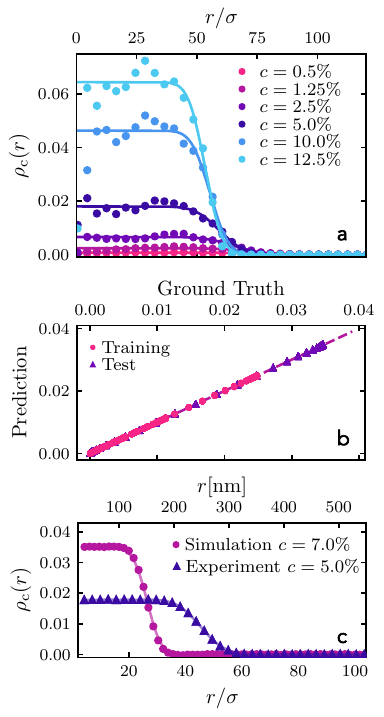}
\end{center}
\vspace{-0.5cm}
\caption{\textbf{Crosslinker distribution prediction} a)  Radial density profiles of the crosslinkers $\rho_c(r)$ for microgels with $N\sim 336000$ at various $c$ from simulations (symbols). Lines corresponds to fuzzy sphere fits (Table S1~\cite{sm}); b) parity plot of target and predicted $\rho_c(r)$ for the training set corresponding to $c=1.25,2.5,5.0,10.0$\% (circles) and test set (triangles); c) predicted $\rho_c(r)$ outside the training set for simulations with $c=7.0\%$ (circles) and experimental data with $c=5.0\%$ triangles) from Ref.~\cite{hazra2023structure}, lines are fuzzy sphere fits. 
}
\label{fig:crosslinker}
\end{figure}

Once we trained the entire AE, we discard the decoder and apply the encoder, which projects the data into the latent space. We report in Fig.~\ref{fig:ae_dens}(d) the probability distributions $P(L)$ of the latent space values $L$ for different $c$ of the training data set with $N\sim336000$. These results confirm that $d'=1$ is already able to distinguish between microgels with different $c$, since the distributions in the latent space are well separated, each following a normal distribution with a maximum standard deviation of $0.01$.  We also observe that the mean values of the distributions, $\bar{L}$, shown in Fig.~\ref{fig:ae_dens}(e), as a function of $c$, are unevenly distributed in the latent space, following a power law relation with $c$ as $\bar{L}(c) \sim A_0 c ^{\nu}$,
where $A_0 \sim 0.514$, and $\nu\sim 0.59$ are fit parameters. 
The latter is remarkably close to the Flory exponent of polymer chains in good solvent. 
A similar value of $\nu$ is recovered by varying the number of neurons in the AE hidden layer. We conclude that the AE is able to capture the underlying physics of the system, with information taken solely from the density profile. Hence, the crosslinker concentration $c$ can be easily predicted for any microgel by inverting  $\bar{L}(c) \sim A_0 c ^{\nu}$ and using the projection on the latent space of a given density profile. 

\begin{figure*}[t]
\begin{center}
\includegraphics[width=0.98\linewidth]{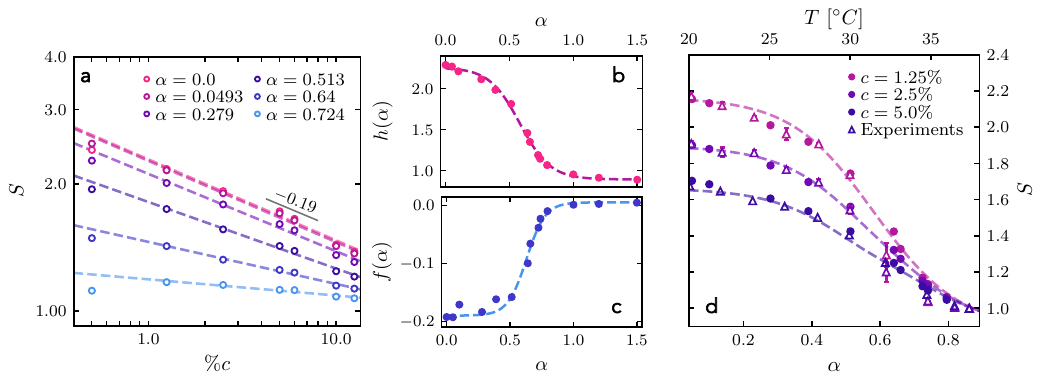}
\end{center}
\vspace{-0.5cm}
\caption{\textbf{Swelling behavior} a) Swelling ratio $S$ as a function of crosslinker molar concentration $c$ for various $\alpha$ values, ranging from $\alpha=0$, the swollen state, up to $\alpha=0.724$. Dashed lines are power-law fits with Eq.~\ref{eq:S-powerlaw}, determining $h(\alpha)$ and $f(\alpha)$, reported in b) and c) as a function of $\alpha$. d) $S$ as a function for $\alpha$ for various values of $c$ from simulations (circles), experimental  experimental data from Ref.~\cite{hazra2023structure} (triangles) and predictions from Eqns.~\ref{eq:S-powerlaw}-\ref{eq:powerlaw} (dashed lines).
}
\label{fig:sw}
\end{figure*}
Indeed, we report in Fig.~\ref{fig:ae_dens}(e) the latent values as a function of the predicted $c$ of all simulated microgels. The AE predictions show remarkable agreement with the expected $c$-values of training and test data sets, with a confidence value ranging from $\pm 1\%$ to $\pm 10\%$. %
We validate the method by applying it to microgels of different sizes. Apart from a small shift,
very close values for the latent space are found for the two considered microgel sizes, as shown in the inset of Fig.~\ref{fig:ae_dens}(e). In addition, the AE is applied to experimental data in Ref.~\cite{hazra2023structure}, providing accurate estimates of $c$, as shown in Fig.~\ref{fig:ae_dens}(e), despite using a single density profile obtained from the fit of the form factor fuzzy sphere~\cite{hazra2023structure}.

Once obtained $c$, we focus on the crosslinkers density profiles $\rho_c(r)$, calculated directly from simulations and shown in Fig.~\ref{fig:crosslinker}(a). Similarly to the total density profiles, they are roughly constant in the core region, although with larger statistical noise due to the small amount of crosslinkers. However, they decay to zero at much shorter distances in $r$ with respect to $\rho(r)$, signaling the accumulation of the crosslinkers within the core. In addition, this decay distance does not strongly depend on $c$ in contrast to what is observed for $\rho(r)$. It is worth noting that also crosslinker profiles are found to be quite well-described by the fuzzy sphere model. The corresponding fits are also shown in Fig.~\ref{fig:crosslinker}(a) with continuous lines with the fit parameters reported in Table S1 of the SM~\cite{sm}. 

We then employ a NN to learn $\rho_c(r)$ from the overall density profile from the same training and test data sets as for the AE. As before, our input data set comes from the discretized total density profile with the $r$-axis normalized. The output now corresponds to the averaged $\rho_{c}$ computed from the fuzzy sphere fits, normalized in the radial axis and in the $y$-axis by the same factor of the total density profile and by the crosslinker concentration $c$, respectively. 
The chosen NN has $1$ hidden layer with $96$ neurons. 
To train the NN, we minimize an error function between the target $\rho_{c}$ and the NN output (see EM). 
The input dataset includes data from microgels assembled with crosslinker concentrations $c=1.25,2.5,5.0,10.0$\%, with additional simulated values of $c$ used as test data sets. In the same spirit of the previous section, we add variability to the input dataset by using short-time-average density profiles as input. 

The parity plot between the ground truth and the predicted $\rho_c(r)$ is reported in Fig.~\ref{fig:crosslinker}(b), where  we observe remarkably good predictions for the training and also for the test data sets. This is confirmed by directly comparing the ML-predicted $\rho_c(r)$ for a $c=7.0$\% microgel outside the training set with the one obtained from the simulations in Fig.~\ref{fig:crosslinker}(c).  
We further test our method by applying it to the $c=5.0$\% experimental data taken from Ref.~\cite{hazra2023structure}, whose ML-predicted $\rho_c(r)$ is also reported in Fig.~\ref{fig:crosslinker}(c). In both cases the predicted profiles are in excellent agreement with those obtained by simulations, showing the robustness of our approach.

Our final goal is to predict the swelling behavior of a microgel, given the ML-predicted crosslinker concentration. To do so, we first simulate the various microgels with different $c$ at many different temperatures, by employing the solvophobic potential $U_{\alpha}$ in Eq.~\ref{eq:alpha}~\cite{gnan2017silico} (see EM). 
We then calculate the swelling ratio $S$ from Eq.~\ref{eq:S}
and report it in Fig.~\ref{fig:sw}(a) as a function of $c$ for different values of $\alpha$ for $N\sim42000$ microgels. 
We then fit the dependence of $S$ on $c$ with a generalized power-law as,
\begin{equation}
    S(\alpha,c)=\frac{R_h}{R_{h0}}=h(\alpha)c^{f(\alpha)},
    \label{eq:S-powerlaw}
\end{equation} 
where $h(\alpha)$ and $f(\alpha)$ solely depend on $\alpha$ and
are reported in Fig.~\ref{fig:sw}(b) and (c), respectively. 
The $\alpha$-dependence of both functions can be written as, 
\begin{equation}
    F(\alpha)\sim A_0^X+A_1^X \zeta((\alpha+A_2^X)/A_3^X),
    \label{eq:powerlaw}
\end{equation}
where $ \zeta(x)=1/(1+e^x)$ is a logistic function, $X$ refers to either $f$ or $h$ and $A_i^X$ are fit parameters associated to each function, reported in Table S2 of the SM~\cite{sm}. We find that the physics of the VPT is directly encoded in these parameters for $f(\alpha)$, being $A_1^f\sim 0.19$ the expected Flory-Rehner exponent at $\alpha=0.0$ and $A_2^f\sim-0.63$, coinciding with the $\alpha$ value at which the VPT occurs. A similar value is also found for $A_2^h$.

Once we know the dependence of $h$ and $f$ on temperature and $c$, we can finally reconstruct the full swelling behavior of any standard pNIPAM microgel. To test this hypothesis, we show the predictions of Eqns.~\ref{eq:S-powerlaw}-\ref{eq:powerlaw} in Fig.~\ref{fig:sw}(d) for different $c$ values in comparison to simulation results and to the experimental data of Ref.~\cite{hazra2023structure}. We find very good agreement at all temperatures, with minor deviations close to the VPT, where experimental error bars are largest. Interestingly, once $S(\alpha)$ is known for example from DLS measurements, it can be used backwards to estimate $c$ via Eq.~\ref{eq:S-powerlaw}, and then an approximate density profile, if not available experimentally, by using our database, following the mapping described in the SM (Section~S4).



In summary, in this work we established a novel ML-approach that, through the sole knowledge of the total density profile of any standard microgel, is able to recover its crosslinker concentration, the crosslinker distribution and the full swelling curve. While the monomer density profile can be directly or indirectly obtained in experiments, 
the crosslinker profiles currently cannot be recovered experimentally. The present study exploits neural networks for the first time to predict them also for experimental systems. While we have validated the approach against the corresponding numerical profiles, it remains a challenge to detect them also experimentally to verify whether they also satisfy a fuzzy sphere decay, as shown in the present simulations. Our work further demonstrates the power of machine learning techniques with carefully selected data coming from simulations to uncover hidden relations. In particular, by using autoencoders (AE), we  are able to extract the information of the crosslinker concentration encoded in the density profiles. The associated latent space is found to follow a power-law dependence on $c$ with an exponent $\sim 0.59$, remarkably close to the Flory prediction for polymers in good solvent, thereby demonstrating the ability of the AE to correctly recover the underlying physics of the system.  
The present methodology can be readily extended to different topologies or to the calculation of other microgel properties, such as elasticity or charge distribution, 
providing sufficient training data. Similarly, it can be applied to assess microgel deformation at interfaces as a function of crosslinker concentration~\cite{camerin2019microgels}. The next step will then be to exploit similar ML-approaches to inverse-design tailored microgel architectures with desired swelling ratios and density profiles to optimize them for specific needs.


We thank L. Tavagnacco for useful discussions. This project was funded by the European Union HORIZON-MSCA-2022-Postdoctoral Fellowships under grant agreement no. 101106848, MGELS. EZ also acknowledges funding from ICSC – Centro Nazionale di Ricerca in High Performance Computing, Big Data and Quantum Computing, funded by European Union – NextGenerationEU - PNRR, Missione 4 Componente 2 Investimento 1.4. We gratefully acknowledge the CINECA award under the ISCRA initiative, for the availability of high-performance computing resources and support.  





\section{End Matter} 
\subsection{Models and methods}

We prepare monomer-resolved microgels using the method  established in Refs.~\cite{gnan2017silico,ninarello2019modeling,del2021two}, which was shown to accurately describe the structure of experimental systems~\cite{ninarello2019modeling,hazra2023structure}. 
After the assembly, we simulate polymer networks with different crosslinker molar fractions $c\in[0.5,1.25,2.0, 2.5, 3.0, 4.0,5.0,6.0,7.0,10.0,12.5,15.0]$, 
covering the full range of experimentally synthesized pNIPAM microgels. We fix the total number of particles in the microgel either to $N\sim 42000$ or $N\sim 336000$ particles, for which $N_c=cN/100$ particles correspond to crosslinkers (particles with four bonded neighbours), with the remaining to monomers (particles with two bonded neighbours).  All particles interact with the Weeks-Chandler-Andersen (WCA) potential, 
\begin{equation}
    U_{WCA}(r)=\begin{cases} 4\epsilon\left[ \left(\frac{\sigma}{r}\right)^{12}-\left(\frac{\sigma}{r}\right)^6 \right]+\epsilon & r\leq 2^{1/6}\sigma \\
    0 & r> 2^{1/6}\sigma,
    \end{cases}
\end{equation}
where $r$ the distance between two beads, $\sigma$ is the diameter of each bead, setting the unit of length, and $\epsilon$ controls the energy scale and sets the unit of energy. For bonded beads, there is an additional interaction modeled by the Finite-Extensible-Nonlinear-Elastic (FENE) potential~\cite{kremer1990dynamics}:
\begin{equation}
    U_{FENE}(r)=-\epsilon k_F {R_0}^2\log{\left[ 1-\left(\frac{r}{R_0\sigma}\right)^2 \right]} \text{ if } r<R_0\sigma,
\end{equation}
where $R_0$ is the maximum bond distance, that is fixed to $R_0=1.5$, and $k_F=15$ controls the stiffness of the bonds. To simulate the effect of temperature, we add a solvophobic attraction~\cite{soddemann2001generic,gnan2017silico}, which mimics the change of affinity of the monomers with respect to the solvent in an implicit way. This potential reads as,
\begin{equation}
    U_{\alpha}(r)=\begin{cases}
    -\epsilon \alpha & r\leq 2^{1/6}\sigma \\
    \frac{1}{2}\alpha\epsilon\left[\cos{\left(\gamma\left(\frac{r}{\sigma}\right)^2+\beta\right)-1}\right] & 2^{1/6}\sigma<r\leq R_0, \\
    0 & r> R_0
    \end{cases}
    \label{eq:alpha}
\end{equation}
where $\gamma=(\pi(2.25-2^{1/3}))^{-1}$, $\beta=2\pi-2.25\gamma$, and $\alpha$ modulates the strength of the attractive interaction, thus amounting to an effective temperature. The network is in good solvent conditions for $\alpha=0$, undergoes the VPT for $\alpha \sim 0.63$ and reaches the collapsed state observed in experiments for $\alpha \sim 0.8-1.0$. Previous studies demonstrated that $\alpha$ varies almost linearly with temperature across the VPT range~\cite{ninarello2019modeling,hazra2023structure}.

We perform NVT MD simulations of individual microgels using the LAMMPS simulation package~\cite{thompson2022lammps} with time step $\delta t=0.002\tau$, where $\tau=\sqrt{m\sigma^2/\epsilon}$ corresponds to the time unit. The temperature is fixed to $k_BT/\epsilon=1.0$, where $k_B$ is the Boltzmann constant, while we vary the solvophobic parameter $\alpha$ to mimic the effect of the real temperature, as described above. The center of mass of the microgel is fixed to the center of the simulation box. We equilibrate the system for at least $3 \times 10^6$ timesteps followed by a production simulation of $1 \times 10^7$ timesteps.
We average results over at least three independent  topologies to take into account the role of disorder in the assembled polymer networks.

The main structural information used as input for the neural networks is the radial density profile of the monomers, with respect to the microgel center of mass, that is averaged over different time intervals. The averaged radial density profile $\rho(r)$ is defined as,
\begin{equation}
    \rho(r)=\left<\frac{1}{N}\sum_{i=1}^N \delta(|\mathbf{r}_i-\mathbf{r}_{cm}|-r)\right>_{\Delta t},
    \label{eq:densprof}
\end{equation}
where $\mathbf{r}_{cm}$ corresponds to the position of the center of mass of the microgel, $\mathbf{r}_i$ that of particle $i$, and $<>$ represent a time average over the window of time $\Delta t$. 
In order to generate a large pool of density profiles and maximize the variance related to noise coming from the simulations, we generate such profiles every $1 \times 10^4$ timesteps averaging every $1 \times 10^5$ timesteps inside the previous time window. In addition, we calculate the radial density profiles of the crosslinker monomers only, $\rho_c(r)$, by substituting $N$ in Eq.~\ref{eq:densprof} with $N_c$. 

We fit the density profiles using the well-established fuzzy sphere model~\cite{stieger2004small}, that is obtained by combining the profile of a solid sphere of radius $R$ with a Gaussian function representing the corona of the particle~\cite{stieger2004small}.  In an approximated version~\cite{scheffold2024revisiting} that holds for standard core-corona microgels, this reduces to 
\begin{equation}
    \rho(r)\sim A  \textrm{ erfc}\left(\frac{r-R_c}{\sqrt{2}\sigma_s}\right),
    \label{eq:fuzzysphere}
\end{equation}
where $A$ is a fit parameter, $R_c$ denotes the radius of the core, and $\sigma_s$ the half-width of the corona shell.
The same model in Eq.~\ref{eq:fuzzysphere} is also used to fit the crosslinker density profiles.  The corresponding fit parameters are reported in the SM~\cite{sm}, Table S1.


We also compute the hydrodynamic radius of the microgel, following the method established in Ref.~\cite{del2021two}, where
\begin{equation}
    R_H=2\left[\int_0^\infty \frac{1}{\sqrt{(a^2+\theta)(b^2+\theta)(c^2+\theta)}}d\theta\right]^{-1},
\end{equation}
 with $a$, $b$, and $c$ the principal semiaxes of the gyration tensor associated to the semplices of the convex hull formed by the microgel particle.  Then, we  calculate the $\alpha$-dependent swelling ratio $S_{\alpha}$, defined as 
\begin{equation}
    S_{\alpha}=R_H(\alpha)/R_H (\alpha^*=0.86),
    \label{eq:S}
\end{equation}
where $R_H (\alpha^*=0.86)$ roughly corresponds to the hydrodynamic radius of the collapsed microgel at the effective temperature $\alpha^*=0.86$. The latter is estimated from  the experimental data of Ref.~\cite{hazra2023structure} and also from Fig.~\ref{fig:sw}(c) where $f(\alpha^*)\sim 0$. 
 
\subsection{Machine Learning Methods}

To learn different structural features of the microgels, we employ two types of machine learning (ML): unsupervised ML with autoencoders (AE)~\cite{tschannen2018recent,boattini2020autonomously} and supervised ML with the use of neural networks (NN)~\cite{gurney2018introduction}. The first method is employed to determine the crosslinker concentration, while the second one is used to calculate the crosslinker distribution.
We use PyTorch~\cite{paszke2019pytorch} to build and train both the AE and the supervised NN.

\subsection{Autoencoders}
An autoencoder is a dimensionality reduction method that consists of a neural network whose main task is to reproduce its input as its output by passing through an intermediate step where the initial input is mapped onto a lower dimension, as explained in the  main text. 
As input, we use a collection of discretized radial density profiles from MD simulations of microgels with different $c$ values, represented by a set of vectors $\mathbf{P}(i)\in \mathbb{R}^d$, where $\mathbf{P}(i)=\mathbf{\rho}_i$ denotes the discretized density profile of sample $i$ and $d$ the dimensionality of the vector. The choice of $d$ allows us to control the resolution of the density profile. Here, we choose $d=115$. In addition, the entries $\mathbf{P}_j(i)$ of the input vector correspond to the value of the density profile evaluated at $r_j$, and thus, we ensure that for all considered data sets, regardless of the crosslinker concentration, the entry $j$ always corresponds to the same distance $r_j$. In order to guarantee this condition and to be able to compare microgels of different sizes, we preprocess the data coming from the radial density profiles as follows. First, we use a scaled radial coordinate $r^*$ that preserves the characteristic decay of each state point. This is done using the value of the integral of the density profile, which in simulations corresponds to the the total number of particles in the microgel $N=\int \rho dV$, and rescaling $r^*=r/N^{1/3}$. 
Second, $\rho(r^*)$ is splined and $P(i)$ is constructed by evaluating the splines in intervals of $r^*=0.025$ starting from $r^*=0.125$. In the output layer, the density profiles, represented by a vector $\mathbf{P}'(i)\in \mathbb{R}^d$, are recovered.  Inside the encoder, we perform a nonlinear projection onto a low-dimensional space  $\mathbf{Y}(i)\in \mathbb{R}^{d'}$, where $d'<d$. We choose $d'$ in such a way that more than $99\%$ of the variance of the input data set is recovered, in this case, corresponding to $d'=1$.  We use a hyperbolic tangent as activation function~\cite{sharma2017activation} varying the number of neurons in the hidden layer from $32$ up to $160$ and the batch size in order to obtain the best parameters that minimize the error function. In particular, we fix the number of neurons in the hidden layer to $80$ , as we do not find relevant changes in the results obtained with different architectures.  
To train the AE, we minimize an error function using mini-batch stochastic gradient descent with momentum~\cite{bishop1995neural,rumelhart1986learning,sutskever2013importance}. In particular, we use the mean squared error function with the addition of a weight decay regularization term~\cite{bishop1995neural} defined as,
\begin{equation}
    E=\frac{1}{M}\sum_{i=1}^{M} \left| \mathbf{P}(i)-\mathbf{P}'(i)\right|^2+\lambda\sum_{j=1}^{W}w_j^2,
    \label{eq:error}
\end{equation}
where $W$ is the total number of weights $w$, $M$ the total number samples in the training data set, and $\lambda=10^{-5}$. In addition, we use a learning rate of $0.01$. We measure the performance of the AE by calculating the fraction of variance explained (FVE) by the NN, defined as,
\begin{equation}
FVE=1-\frac{\text{MSE}}{\text{var}(P)}=1-\frac{\sum_{i=1}^M \left| \mathbf{P}(i)-\mathbf{P'}(i)\right|^2 } {\sum_{i=1}^M {\left| \mathbf{P}(i),-\left<\mathbf{P}\right>\right|^2} } 
\end{equation}
where the MSE corresponds to the Mean-Squared-Error and $\text{var}$ denotes the variance of $\mathbf{P}$.

\subsection{Neural Networks}
To predict the distribution of the crosslinkers within the microgel, we use simple NN with one hidden layer and ReLU as the activation function~\cite{sharma2017activation}. The same normalized discretized radial density profiles $\mathbf{P}(i)\in\mathbb{R}^d$ with $d=115$ from the AE are used as input to the NN. In this case, the feature to learn corresponds to the crosslinker distribution $\rho_c$ represented by the vector $P'$. Since this quantity is subject to large statistical noise due to the small number of involved crosslinkers, we use the  fuzzy sphere fits of $\rho_c$, averaged over different microgel configurations, for each $c$. As for the AE, the radial coordinate is normalized with the total number of particles $1/N^3$. Due to the small values of $\rho_{c}$, the $y$-axis is also normalized by the corresponding value of $c$. 
The number of neurons in the hidden layer is varied from $32$ to $736$. We minimize the error function defined in Eq.~\ref{eq:error} between the predicted crosslinker density profiles $\mathbf{P'}$ and the original ones from the training set, using mini-batch stochastic gradient descent with momentum~\cite{bishop1995neural,rumelhart1986learning,sutskever2013importance} and a learning rate of $0.1$.
The NN is trained until the number of epochs reaches $5 \times 10^6$ or the $MSE$ difference between the last $500$ epochs is less than $1 \times 10^{-7}$. We finally fix the number of neurons to $96$ reaching a $RMSE=1.3 \times 10^{-6}$. 

\clearpage
\newpage
\onecolumngrid
\begin{center}
\large
\textbf{Predicting structure and swelling of microgels with different crosslinker concentrations combining machine-learning with numerical simulations\\ \bigskip Supplementary Material}

\normalsize
\bigskip
Susana Marín-Aguilar\textsuperscript{ 1}, Emanuela Zaccarelli\textsuperscript{ 1,2}\\
\medskip
\small
\textit{%
\textsuperscript{1}Department of Physics, Sapienza University of Rome, Piazzale Aldo Moro, 5, 00185, Roma, Italy\\
\textsuperscript{2}CNR Institute of Complex Systems, Uos Sapienza,Piazzale Aldo Moro, 5, 00185, Roma, Italy \\
}

\end{center}
\onecolumngrid
\renewcommand{\theequation}{S\arabic{equation}}\setcounter{equation}{0}
\renewcommand{\thefigure}{S\arabic{figure}}\setcounter{figure}{0}
\renewcommand{\thetable}{S\arabic{table}}\setcounter{table}{0}
\renewcommand{\thesection}{S\arabic{section}}\setcounter{section}{0}

\section{Scaling of characteristic lengths of the microgels with crosslinker concentration}

\begin{figure}[h]
\begin{center}
\includegraphics[width=0.4\linewidth]{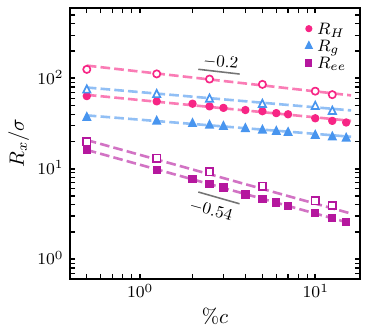}
\end{center}
\caption{Crosslinker concentration $c$ dependence of characteristic microgel lengths ($R_x$): hydrodynamic radius $R_H$ (circles),  gyration radius $R_g$ (triangles) and end-to-end distance $R_{ee}$ (squares) for microgels with $N\sim 42000$ (fill symbols) and $N\sim 336000$ (open symbols). The first two quantities follow a power-law behavior with exponent $\sim -0.216\pm0.007$, while the third one obeys a different power-law with exponent $-0.54 \pm 0.01$. Dashed lines are the corresponding power-law fits.
}
\label{fig:scaling}
\end{figure}

Here we report additional data, already known from previous works and from theoretical considerations, to further validate the present simulations in the swollen conditions $(\alpha=0)$. It is indeed well-established that the crosslinker concentration $c$ has a direct effect on the microgel size, but it is worth reminding that it affects characteristic length-scales in a different way. On the scale of the whole microgel, both the hydrodynamic radius $R_H$ and the gyration radius $R_g$ are known to follow the Flory-Rehner theory with a power-law with $c$, $R_X\sim c^{-0.2}$\cite{hazra2023structure}. This is reported in Fig.~\ref{fig:scaling}, where results from simulations of microgels of both $N\sim 42000$ and $N\sim 336000$ as a function of $c$ are shown. However, considering variations on a more local scale, as for example 
the end-to-end distance $R_{ee}$ of chains composing the network, which are also reported in Fig.~\ref{fig:scaling}, a different exponent is observed. We find $R_{ee} \sim c^{-0.54}$, that is reasonably close to the $\nu$ exponent from the Flory theory of polymers under good solvent conditions. The observed small deviation may be due to the presence of the network which slightly constrain the behavior of the chains.
Importantly, both scaling behaviors and related exponents are essentially independent of the microgel size, as expected.

\section{Fuzzy sphere model fits of density profiles of the whole microgel and of the crosslinkers only}

The total density profiles $\rho(r)$, shown in Fig.~1(a) of the main text for microgels with $N\sim42000$ and different crosslinkers at $\alpha=0$ are accurately described with the fuzzy sphere model defined in Eq.~7 in the main text described as,
\begin{equation}
    \rho(r)\sim A  \textrm{ erfc}\left(\frac{r-R_c}{\sqrt{2}\sigma_s}\right).
    \label{eq:fuzzysphere}
\end{equation}
The corresponding fit parameters are reported in Table~\ref{tab:fuzzyfitdensprof}, also for the larger simulated microgels ($N\sim 336000$).
Similarly, we find that also the crosslinker density profiles $\rho_c(r)$, shown in  Fig.~2(a), are well captured by the same model, displaying much smaller coronas with respect to the overall density profiles. The corresponding fit parameters for both microgel sizes at different $c$ are displayed in Table~\ref{tab:fuzzyfitdensprof}. Note that for small values of $c$ for $N\sim 42000$, we do not report fit parameters due to the very low number of crosslinkers, which make $\rho_c$ to exhibit large statistical deviations, resulting in noisy fits.

\begin{center}

\begin{table}[ht!]
\centering
\renewcommand{\arraystretch}{0.6} 
\newcolumntype{C}[1]{%
 >{\vbox to 1ex\bgroup\vfill\centering\arraybackslash}%
 p{#1}%
 <{\vskip-\baselineskip\vfill\egroup}}  

\begin{tabular}{|C{1cm}|C{1cm}|C{1cm}|C{1cm}|C{0.2cm}|C{1cm}|C{1cm}|C{1cm}|}
\hline 

& \multicolumn{3}{|C{3cm}|}{Overall $\rho(r)$} & &\multicolumn{3}{|C{3cm}|}{$\rho_c(r)$} \\ 
\hline

\hline
$\%c$ & $A $ & $R_c$ & $\sigma_s$ & & $A $ & $R_c$ & $\sigma_s$ \\ 
\hline \hline
\multicolumn{8}{|C{7.2cm}|}{$N\sim 42000$} \\ 
\hline
$0.5$ & $0.061$ & $40.0$ & $9.9$ &  & $//$ &  $//$ &  $//$  \\
$1.25$ & $0.087$ & $35.4$ & $10.8$& & $//$ &  $//$ &  $//$   \\
$2.5$ & $0.112$ & $33.0$ & $8.56$ & &  $//$ &  $//$ &  $//$  \\
$5.0$ & $0.152$ & $30.3$ & $6.66$ & &  $//$ &  $//$ &  $//$ \\
$7.0$ & $0.170$ & $29.58$ & $5.56$ & & $0.016$ & $26.92$ & $3.48$   \\
$10.0$ & $0.198$ & $28.0$ & $4.37$ & &$0.026$ & $26.44$ & $3.32$   \\
$12.5$ & $0.215$ & $28.0$ & $3.77$  & &$0.035$ & $25.77$ & $3.55$   \\
 \hline
 \hline
 \multicolumn{8}{|C{7.2cm}|}{$N\sim 336000$} \\ 
\hline

$0.5$ & $0.048$ & $89.94$ & $14.9$ &  & $0.0003$ & $77.73$ & $12.84$ \\
$1.25$ & $0.071$ & $79.37$ & $15.13$&  & $0.001$ & $71.95$ & $10.62$\\
$2.5$ & $0.101$ & $70.80$ & $13.34$ & & $0.003$ & $65.37$ & $9.47$ \\
$5.0$ & $0.149$ & $62.64$ & $10.73$  & & $0.009$ & $59.4$ & $8.01$  \\
$10.0$ & $0.160$ & $63.52$ & $8.31$ & & $0.023$ & $55.31$ & $6.01$ \\
$12.5$ & $0.230$ & $55.45$ & $6.05$  & & $0.035$ & $53.45$ & $5.52$  \\
 \hline
\end{tabular}
\caption{Fuzzy sphere~\cite{bergmann2018super,ninarello2019modeling} fitting parameters from Eq.~\ref{eq:fuzzysphere} of radial density profiles $\rho(t)$ and crosslinker density profiles $\rho_c(t)$ coming from microgels with $N\sim 42000$ and $N\sim336000$ and various crosslinker concentrations $\%c$, where $A$ a fitting parameter, $R$ corresponds to the radius of the core, and $\sigma$ the width of the corona. }
\label{tab:fuzzyfitdensprof}
\end{table}
\end{center}

\section{Prediction of swelling behavior}

In the manuscript we showed that we can reach a full knowledge of the entire swelling curve of the microgels as a function of crosslinker concentration, by phenomenologically fitting the auxiliary functions $f(\alpha)$ and $h(\alpha)$ with a logistic function (Eq.~2 of the main text). The corresponding fit parameters are reported in 
Table~\ref{tab:swelling}.

\begin{center}

\begin{table}[h!]
\centering
\renewcommand{\arraystretch}{0.6} 
\newcolumntype{C}[1]{%
 >{\vbox to 1ex\bgroup\vfill\centering\arraybackslash}%
 p{#1}%
 <{\vskip-\baselineskip\vfill\egroup}}  

\begin{tabular}{|C{0.8cm}|C{1.4cm}|C{1.4cm}|C{1.4cm}|C{1.4cm}|}
\hline 
$ $ & $A_0 $ & $A_1$ & $A_2$ & $A_3$  \\ 
\hline \hline
$h(\alpha)$ & $2.26$ &  $-1.37$ & $-0.58$ &  $0.12$ 
\\
$f(\alpha)$ & $-0.19$ & $0.19$ &$-0.63$ &  $0.074$  \\
 \hline
\end{tabular}
\caption{Fit parameters $A_0$, $A_1$, $A_2$, $A_3$ from Eq.~2 for $h(\alpha)$, which corresponds to the amplitude function of the swelling ratio in Eq.~1 of the main text, and $f(\alpha)$ that is the corresponding power-law exponent, both varying with $\alpha$. We note that $A_2$ is very close for both functions, expressing the value of $\alpha$ at the VPT. }
\label{tab:swelling}
\end{table}
\end{center}

\section{Prediction of experimental density profiles and scaling of the $r$-coordinate to experimental values}

\begin{figure}[h]
\begin{center}
\includegraphics[width=0.9\linewidth]{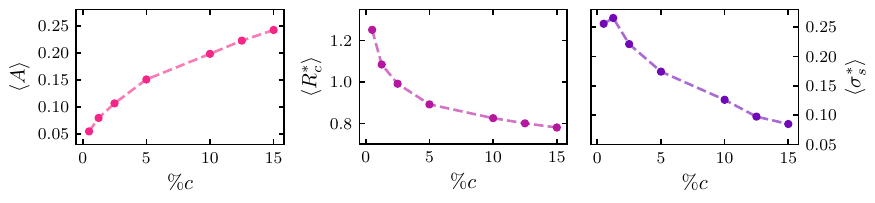}
\end{center}
\caption{Averaged normalized fuzzy sphere fit parameters a) $\left<A\right>$, b) $\left<R_c^*\right>$ corresponding to the normalized radius of the core and c) $\left<\sigma^*_s\right>$ the normalized half-width of the shell as a function of crosslinker concentration $c$. Both lengths are relative to the normalized distance $r^*$ and should be rescaled using Eq.~\ref{eq:rexp} to be converted to experimental units. Instead, the $A$ parameter should be used as it is. Lines are guides to the eye. 
}
\label{fig:normalized_fuzzy}
\end{figure}

Based on the extensive information from simulations, provided in the manuscript, with the full dependence of density profiles on crosslinker concentration $c$, it is also possible to make an estimate of the experimental density profile at low temperatures based on $c$ only. To do so, we estimate the fuzzy sphere fit parameters of the normalized (by using $r^*$) density profiles from $N\sim42000$ and $N\sim 336000$ and average their values. We show  the dependence of these values on $c$ in Fig.~\ref{fig:normalized_fuzzy}. Once the fitting parameters are obtained as a function of $c$, an estimate of the density profiles as a function of $r^*$ can be obtained for each $c$ by using the (approximate) fuzzy sphere model in Eq.~7. Then, in order to  rescale the $x$-axis,
we use the mapping established in Ref.~\cite{hazra2023structure} between experiments and simulations. In brief, for a microgel with $R_H\sim 300$ nm, the simulation bead size for a $N\sim336000$ microgel used in our simulations is 3.34nm. This value should be used as a reference, therefore depending on the size of the  microgel,
the correct factor to be used for estimating the experimental density profiles, both the total one and the crosslinker one, is:
\begin{equation}
r_{exp}=r^* \left(\frac{3.34}{300}\right) R_H \ \text{nm},
\label{eq:rexp}
\end{equation}
with $R_H$ the experimental value of the hydrodynamic radius to be expressed in nm.

\clearpage
\newpage
\twocolumngrid

\bibliography{prl_arxiv}
\bibliographystyle{apsrev4-2}

\end{document}